\documentclass[12pt,a4paper]{article} 
\usepackage{calc,amsmath,amsthm,amssymb,color,graphics,natbib}
\usepackage{appendix}
\usepackage{fullpage}
\usepackage{fancyhdr}
\usepackage{standardmacros}
\usepackage{rotating}
\usepackage[margin=6mm]{caption} 

\newdimen\np
\title{\vspace*{-1.8cm}Conjugate distributions in hierarchical Bayesian ANOVA
for computational efficiency and assessments of both practical and statistical
significance}
\author{
\begin{minipage}{0.5\textwidth}
\begin{center}
Steven Geinitz \\
\begin{normalsize}
\textit{geinitz@math.uzh.ch} \\
University of Zurich \\
\vspace{-1mm}
Zurich, Switzerland 
\end{normalsize}
\end{center}
\end{minipage}
\begin{minipage}{0.5\textwidth}
\begin{center}
Reinhard Furrer \\
\begin{normalsize}
\textit{furrer@math.uzh.ch} \\
University of Zurich \\
\vspace{-1mm}
Zurich, Switzerland 
\end{normalsize}
\end{center}
\end{minipage}
}
\date{}
\begin{document}
\maketitle
\begin{abstract}  
\vspace{-1cm}
\noindent
\textbf{Abstract}
Assessing variability according to distinct factors in data is a fundamental
technique of statistics.  The method commonly regarded to as analysis of
variance (ANOVA) is, however, typically confined to the case where all levels of
a factor are present in the data (i.e.\ the population of factor levels has been
exhausted).  Random and mixed effects models are used for more elaborate cases,
but require distinct nomenclature, concepts and theory, as well as distinct
inferential procedures.  Following a hierarchical Bayesian approach, a
comprehensive ANOVA framework is shown, which unifies the above statistical
models, emphasizes practical rather than statistical significance, addresses
issues of parameter identifiability for random effects, and provides
straightforward computational procedures for inferential steps.  Although this
is done in a rigorous manner the contents herein can be seen as ideological in
supporting a shift in the approach taken towards analysis of variance. \\ \\
\noindent
\textit{Keywords}:\, ANOVA; fixed effects; random effects; variance components; 
hierarchical Bayes; multilevel model; constraints
\end{abstract}

\section{Introduction}
As an independent field of study Statistics is rather young.  Many of the
methods, techniques, and philosophies can be attributed to a handful of
statisticians during the first half of the twentieth century.  Among these, R.A.
Fisher is often recognized as having had a profound influence on the field.  It
has been said that Fisher single-handedly created the foundations of modern
statistical science \citep{Hald:98}.   For statisticians the first
contribution that comes to mind is his work in development of likelihood theory.
However, for the greater scientific community, one might consider his
formulation of analysis of variance as the most significant contribution.  \\ \\
As much of Fisher's work was in agriculture, an apt example to consider is the
one-way ANOVA, $Y_{ij} = \mu + \alpha_i + \epsilon_{ij}$, in which observations
are on crop yield, with $i,j$ representing the $j^{th}$ plant receiving
fertilizer treatment $i$.  An appropriate decomposition of the data should then
reveal the variability due to different fertilizers while accounting for
variability within plant groups that receive the same type of fertilizer
treatment.  Thus, analysis of variance is essentially a pragmatic decomposition
of the data.  In correspondence Fisher has been cited \citep{Sear:etal:92} to
have said,
\begin{quote} 
  ``The analysis of variance is (not a mathematical theorem but) a simple method
  of arranging arithmetical facts so as to isolate and display the essential
  features of a body of data with the utmost simplicity.''  
\end{quote}
The elegance and power of this methodology is perhaps what has caused ANOVA to
become so popular in nearly all areas of scientific research.  However, along
with the ubiquitous support of the methodology has come a pervasive reliance on
its conclusory result, the $p$-value.  Recognition of this problem is not new.
It has been long noted by researchers in other fields that the hypothesis-based
point of view, which relies on statistical significance, should be amended.
\citep{Yocc:91, Fidl:etal:04, Ioan:05}.  The statistical community has also long
acknowledged the need to provide methodologies that are first and foremost, ``of
use to scientists in making quantitative inferences,'' \citep{Neld:99}.  The
problem is that the standard methods that continue to be imparted on students
focus on statistical significance.  As stated by \citet{Sava:57}, a method that
does so ``simply reflects the size of the sample and the power of the test, and
is not a contribution to science.''  Thus, any standard, or default methodology
that aims to decompose variation present in a set of observations according to
factors of interest, should be able to address practical significance as well.  

In addition to the base objective of analysis of variance, to decompose
variation in observations according to distinct sources of variability, a
default method used in initial/exploratory work should accomplish the following. 
\begin{itemize}
  \item Allow for each factor to simultaneously consider variability due to the
    observed set of effects (finite population variance), as well as the
    variability from unobserved effects (superpopulation variance), thereby
    permitting greater flexibility in model choice with regards to fixed or random
    effects.
  \item Facilitate comparison of magnitude of variability across all factors in
    the model, including errors, so that attention may be given to practical
    significance as well as statistical significance of a factor.
  \item Provide ability to consider both magnitude and uncertainty of variance
    parameters in the model, by providing confidence, or uncertainty intervals
    in a default analysis summary.
\end{itemize}
These are precisely the goals of the analysis of variance framework proposed in
this paper.  While the primary contribution may be seen as ideological in
nature, there are technical issues that are addressed to allow for a shift in
the standard approach taken towards the basic method of analysis of variance.
By standard approach one may assume the tabular analysis of variance summary and
its accompanying test of statistical significance.

The organization of the paper is as follows.  Section~\ref{sec:backgrnd} covers
basic concepts of standard methods that are both widely taught and employed, as
well as recent shifts in the practice of ANOVA.  Section~\ref{sec:comprANOVA}
presents an alternative framework of ANOVA along with modifications to the
standard ANOVA table summary.  Section~\ref{sec:examples} illustrates our method
and compares it to the classical approaches.  In particular, we present an
example in which classical ANOVA yields identical $p$-values for two cases; one
in which the factor under investigation has low practical significance, and one
with high practical significance.  

\section{Background}\label{sec:backgrnd}
Following Fisher's analysis of variance overall uncertainty is attributed to
distinct factors of an experiment through the use of a sum of squares
decomposition.  This is now shown with the balanced one-way analysis of variance
model
\begin{align}
  Y_{ij} = \mu + \alpha_i + \epsilon_{ij}, \qquad i = 1,\dots,n_I, \quad j =
  1,\dots,n_J.\label{eq:1waystandard}
\end{align}
As a seminal example consider observations that are on crop yield with $i,j$
representing the $j^{th}$ plant receiving fertilizer treatment $i$.  More
generally the indices represent a factor level $i$ and replicate $j$.  The
appropriate decomposition of the data, which reveals variability due to
different fertilizers while accounting for variability within plant groups that
receive the same type of fertilizer treatment, is done with the arithmetical
arrangement that summarizes yield for each type, $\overline{Y}_{i.} =
n_J^{-1}\sum_j Y_{ij}$, and for overall yield, $\overline{Y}_{..} = n^{-1}
\sum_i \sum_j Y_{ij} = n_I^{-1} \sum_i \overline{Y}_{i.}$, where $n = n_I \cdot
n_J$.  Observations $Y_{ij}$ are decomposed with the identity
\begin{align}
  Y_{ij} - \overline{Y}_{..} =  ( Y_{ij} - \overline{Y}_{i.} ) + ( \overline{Y}_{i.} -
  \overline{Y}_{..}).\label{eq:ydecomposed}
\end{align}
Terms are then squared and summed, noting that the cross term on the right hand
side equals zero, so that a decomposition of the mean-adjusted sums of squares
is
\begin{align}
  \underbrace{\sum_{i,j} (Y_{ij} - \overline{Y}_{..})^2}_{\SST} & = 
  \underbrace{\sum_{i,j} ( Y_{ij} - \overline{Y}_{i.} )^2}_{\SSE} +
  n_J \underbrace{\sum_{i} ( \overline{Y}_{i.} - \overline{Y}_{..})^2}_{\SSA},\label{eq:sumsq}
\end{align}
where $\overline{Y}_{i.}$ is the mean within group $i$ and $\overline{Y}_{..}$ is the mean
of all observations.  The terms $\SST$, $\SSA$, and $\SSE$ denote  \textit{total
(adjusted)} sum of squares, sum of squares \textit{among} groups, and sum of
squared \textit{errors}, respectively.  Note that each of these terms is itself
a sum of squares that is analogous to a sample variance $s^2 =
k^{-1}\sum_{i=1}^k (x_i - \overline{x})^2$, for a set of independent observations
$x_1, \dots, x_k$, and is thus proportional to a $\chi^2$ distribution with
appropriate degrees of freedom.  Fisher showed that $\SSA$ and $\SSE$ are both
proportional to $\chi^2$ distributions, with $n_I-1$ and $n - n_I$ degrees of
freedom, respectively, and that they are independent, the general result of
which is due to \citet{Coch:34}.  

While this classical methodology provides a means to examine statistical
significance, it does not provide any formal assessment of practical
significance.  Loosely speaking, practical significance can be considered as a
contextual basis that allows data-specific conclusions to be drawn, i.e.\
evidence that $\SSA$ is substantial compared, not only to zero, but to $\SSE$ as
well.  Practical significance in the example above implies that the variability
due to the fertilizer treatment is not only significantly different than no
treatment, but that when compared to plant-to-plant variability it is still
significant.   One contribution of this paper is in attempting to formalize a
statistical methodology that rigorously provides a method of assessing practical
significance.

\subsection{Conventional Methods}                     
A fixed effects model generally refers to the case when the observations have
exhausted the population of factor levels (e.g.\ treatments), or when interest
lies only with the factor levels that have been observed.  Alternatively, random
effects models are employed when it is assumed that the factor levels are a
subset of a greater population of possible levels.  This definition provided by
\citet[p.195]{Hoag:Most:Tuke:91} is somewhat more explicit than that given by
\citet{Eise:47}, in which the effects of a model are considered to be fixed when
they are all nonrandom, and considered to be random when they are all random.
There exist many other definitions in the literature, some of which are
not compatible.  See \citet{Gelm:05a} for a summary.  

\subsubsection{Fixed Effects} \label{ch1:fixedeff}
Consider the model given by \eqref{eq:1waystandard} such that $i = 1,\dots,n_I$
denotes the factor level or treatment, and $j = 1,\dots,n_J$ denotes
replications or errors.  Observations are assumed to be independent across
replicates as well as across factor levels.  Additionally, it is generally
assumed that
\begin{align}
  \epsilon_{ij} \sim N(0, \; \sigma^2_\epsilon). \label{eq:assump1}
\end{align}
Analysis of variance generally aims to test the hypothesis that there is no
difference among the treatments, 
\begin{align}
  H_0: \alpha_1 = \dots = \alpha_{n_I} = 0, \label{eq:hyp1}
\end{align}
against the alternative hypothesis that at least one treatment level differs.
The test is a result of the sums of squares decomposition in \eqref{eq:sumsq},
since $\frac{\SSE}{\sigma^2_\epsilon} \sim \chi^2_{n-n_I}$ and (under the null
hypothesis) $\frac{\SSA}{\sigma^2_\epsilon} \sim \chi^2_{n_I-1}$, where $n =
n_I\cdot n_J$.  The expectation of these two terms is $\frac{n_J}{n_I-1}\sum_i
\alpha_i^2 + \sigma^2_\epsilon$ and $\sigma^2_\epsilon$, respectively.    The
test of $H_0$ is then carried out using the $F$ distributed ratio
$\frac{\MSA}{\MSE}$,
where $\MSA = \frac{\SSA}{n_I-1}$ and $\MSE = \frac{\SSE}{n-n_I}$.  The term $\MSA$ is central
$\chi^2_{n_I-1}$ distributed when \eqref{eq:hyp1} is true, and non-central with
shift of $\frac{n_J}{n_I-1}\sum_i \alpha_i^2 + \sigma^2_\epsilon$ when false.

The results described are concisely displayed in a tabular format
\citep{Fish:25}, as seen in Table~\ref{tab:oneway}.  The table culminates with
\eqref{eq:hyp1} being tested based on the $p$-value of $p =
\textrm{Pr}(F_{n_I-1,n-n_I} > F)$, which does not give any indication of the
practical significance.  And despite recognition of the need to focus on effect
sizes and confidence intervals \citep{Gard:Altm:86, Naka:Cuth:07} rather than
testing, the table remains a staple among statistical methodologies.  
\begin{table}[h]
\begin{center}
\caption{One-way analysis of variance.}
\begin{tabular}{lrrrrr}
 \hline
 Source   & Df                 & Sum Sq & Mean Sq & F value & Pr($>$F) \\ 
 \hline
Factor A  & $n_I-1$      & $\SSA$  & $\MSA$   & $F=\frac{\MSA}{\MSE}$ & Pr$(F_{n_I-1,n-n_I}>F)$ \\ 
  Errors  & $n-n_I$ & $\SSE$  & $\MSE$   &         &  \\ 
   \hline
\end{tabular}
\label{tab:oneway}
\end{center}
\end{table}
 
\subsubsection{Random Effects} \label{sec:randeff}
In addition to the statistical model \eqref{eq:1waystandard} and distributional
assumption \eqref{eq:assump1}, there is an additional assumption on the factor
levels,
\begin{align}
  \alpha_{i} \sim N(0, \; \sigma^2_\alpha), \quad i = 1, \dots n_I, \label{eq:assump2}
\end{align}
with $\alpha_i$ and $\epsilon_{ij}$ independent.  Observations are then normally
distributed with mean and variance
\begin{align*}
  \E[Y_{ij}] = \mu, & \qquad \qquad
  \textrm{Cov}(Y_{ij}, Y_{i^\prime j^\prime}) = 
  \begin{cases}
    \sigma^2_\alpha + \sigma^2_\epsilon & i = i^\prime, j = j^\prime, \\
    \sigma^2_\alpha & i = i^\prime, j \neq j^\prime, \\
    0 & i \neq i^\prime.
  \end{cases}
\end{align*}
This parameterization has the added benefit that the parameter space for the
factor levels is reduced from $n_I$ to $1$, since only $\sigma^2_\alpha$ is
estimated. Although individual levels, $\alpha_i$, may be predicted if
necessary.  Averaging over replications at factor level $i$ yields the mean
$\overline{Y}_{i\cdot}$, which are independently distributed $N(\mu,
\sigma_{\alpha\epsilon}^2)$, where $\sigma_{\alpha\epsilon}^2 = \sigma_\alpha^2 +
\frac{\sigma_\epsilon^2}{n_J}$.  Thus, the likelihood is a function of the three
parameters $\mu,\sigma^2_\epsilon$, and $\sigma^2_\alpha$.   


Analogous to \eqref{eq:hyp1}, the initial inquiry of interest is generally
concerned with whether greater population variance $\sigma_\alpha^2$ is
significantly different from zero.  This corresponds to the null hypothesis
\begin{align}
  H_0: \sigma^2_\alpha = 0, \label{eq:hyp2}
\end{align}
and is tested using the same F-statistic as for \eqref{eq:hyp1}
\citep{Sear:etal:92, Rao:97, Cox:Solo:03}.  Aside from its unintuitive nature,
in that despite being random vs.\ fixed the same test statistic is used, this
hypothesis test does little to remark on the practical significance of the
variation due to factor $\alpha$.  Namely, the hypothesis may be rejected even
when variation due to the errors is substantially greater, as seen in the
example of Section~\ref{subsec:simdata}.

Further inferential procedures on the variance components themselves are
typically based on method of moments estimators, or explicitly use the
likelihood.  In the latter case, variability of the variance components are
estimated with the Hessian of the likelihood, as with the widely used \texttt{R}
packages \texttt{nlme} \citep{Pinh:etal:06} and \texttt{lme4}
\citep{Bate:DebR:04}.  Wald-type confidence intervals can be then used to obtain
confidence regions for the parameters.  Similarly, the asymptotic properties of
the log likelihood can be utilized to obtain confidence intervals using the
$\chi^2$ distribution, as seen in Figure~\ref{fig:ex01:freqvsbanova} of Section
\ref{sec:rails}.

\subsubsection{Issues and Concerns} \label{subsec:issues}
The choice to use a fixed or random effects model is not always immediately
clear.  The terminology alone may be seen as ambiguous since the distinction
between fixed effects, random effects, and mixed effects is somewhat malleable.
The simple fixed effects model of Section~\ref{ch1:fixedeff} can be seen as
having a random component in the errors, $\epsilon_{ij}$.  Similarly, the random
effects model of Section~\ref{sec:randeff} can be seen as having a fixed
component, $\mu$.  In both cases implying a mixed effects model.  In practice a
mixed effects model is employed when there are two or more factors, other than
overall mean and errors, and they are not all fixed (random).  

More difficult perhaps is determining when which of these methods should be
used.  If interest lies in the distribution of the random effects, i.e. the variance
component $\sigma^2_\alpha$, then a random effects model should be chosen.  If
interest lies in the realized/observed levels of the factor, then a fixed
effects model is used.  If both are of interest, then the random effects should
be chosen and levels are then predicted, rather than estimated.
\citet[p18]{Sear:etal:92} take a pragmatic approach to this by recommending that
in any case in which it is reasonable to assume that the levels of the factor
come from a probability distribution, i.e.\ that \eqref{eq:assump2} may be
assumed, then a random effects model should be chosen.  The usage of a random
effects model, however, typically precludes the estimation of the
finite population variance.  

An additional problem that arises in analysis of variance with several factors
is the so-called 'mixed models controversy' \citep{Voss:99, Lenc:etal:05,
Neld:08}.  The problem essentially comes down to how a hypothesis test of a
random effect is carried out when an interaction is also present in the model.

To resolve the issues above we support the notion of \citet{Gelm:05a}, in that
\textit{all} factors in the model are treated as random.  The procedural steps
are then carried out equivalently.  If interest is in the observed (unobserved)
levels of a factor, then the greater focus is given to the finite (super)
population variance.  However, because of parameter dependencies involved in the
unconstrained factor levels, Gelman recommends using MCMC, in which redundant
parameterization is used in order to reduce dependencies and to speed up
posterior sampling.  Alternatively, we recommend using constraints to define an
improper joint prior distribution on the factor levels, thereby eliminating the
need for complex MCMC procedures, as in \citet{Gein:etal:12a}.  


\subsection{Multilevel Models}\label{sec:ch1multilevel}
Often times the results of an analysis should allow for simultaneous
consideration of both group level and individual level variability, e.g.\
variability according to schools and to students within schools.
Applications of such scenarios initially arose in the social sciences
\citep{Gold:95, Kref:deLe:98, Snij:Bosk:11}, but have also included the health
sciences \citep{VonK:etal:92, Gree:00}, and have provided the basis for much of
the work in multilevel models.  

A multilevel model can be seen as a linear model with coefficients, i.e.\ factor
levels, that are themselves modeled \citep{Gelm:06}.  More generally, this can
be considered as a type of hierarchical Bayesian approach.  However, while not
explicit, the multilevel point of view is useful in considering a generalized
approach towards analysis of variance.   Because the simultaneous consideration
of group and individual level variability entails the decomposition of variation
according to each of these sources,``ANOVA is fundamentally about multilevel
modeling'' \citep{Gelm:05a}.  That is to say, analysis of variance from the
viewpoint of multilevel models allows for both finite population and
superpopulation variance components to be considered, which can be seen as a
unification of fixed and random effects.   This comprehensive approach to
analysis of variance yields useful results and has been used in other fields
such as ecology \citep{Qian:Shen:07}, genetics \citep{Lein:08}, and climate
\citep{Sain:etal:11}. 

In practice there have been some hindrances in the adoption of this more general
approach to ANOVA.  Computational procedures to carry out such an analysis
typically rely on either mixed effects models (e.g.\ \texttt{lme4} package in
\texttt{R}) or on MCMC methods (e.g.\ \texttt{WinBUGS}).  However, while mixed
effects models can be used to obtain initial estimates of the parameters in a
multilevel model, inferential steps, e.g.\ confidence intervals, for variance
parameters are often done through likelihood approximation.  For more explicit
inferential procedures it is necessary to use MCMC methods
\citep[p.566]{Gelm:06}.  Although the added complexity and computation of MCMC,
particularly when the use is as an exploratory analysis step, can be a deterrent
to this approach.  A method that is both precise in its inferential statements
while being straightforward to implement is not widely known.

\subsection{Bayesian Results}
The hierarchical approach towards analysis of variance can be
explained most readily in a Bayesian framework.  In an effort to explain this
approach in a classical inference framework \citet{Gelm:05a} recommends a
simulation approach, which is reminiscent of posterior sampling.  Because we
prefer to adopt an explicit Bayesian approach, we now review some results on
distributions for variance components that facilitate the procedure.  

\subsubsection{Prior Distributions}
For the normally distributed random variable $Y \sim \Norm(\mu, \, \sigma^2)$,
prior specification of the parameters can be done in many different ways.
Initially, consider $\mu, \sigma^2$ to be either known or unknown, each in turn.
Following the invariance principle \citep{Jeff:46}, prior distributions in
univariate cases are then
\begin{center}
\begin{tabular}{l r c l}
  $\mu$ known, $\sigma^2$ unknown: & $p( \mu )$ & $\propto$ & const, \\ 
  $\mu$ unknown, $\sigma^2$ known: & $p( \sigma^2 )$ & $\propto$ & $(\sigma^2)^{-1}$.
\end{tabular}
\end{center}
\citet[p.43]{Box:Tiao:92} derive same priors using the concept of location and
scale parameters.  These identical priors are also found using the reference
approach of \citet[p314]{Bern:Smit:94}, due to asymptotic normality of the
posterior distributions.  Note that the density of $p(\sigma^2) = (\sigma^2)^{-1}$
corresponds to an inverse-gamma distribution, $\Gamma^{-1}(u,v)$, with $u = v =
0$.  Common values of hyperparameters have thus been limiting forms thereof,
such as $u = v = \varepsilon$, with $\varepsilon$ small \citep{Lunn:etal:00}.
If prior independence between $\mu$ and $\sigma^2$ is assumed, then the two
univariate priors are combined for 
\begin{align}
  p(\mu, \sigma^2 ) & = p(\mu)p(\sigma^2) \propto (\sigma^2)^{-1} \label{eq:jindpri}.
\end{align}
Alternatively, Jeffreys' prior for multivariate parameters 
$\btheta = (\mu, \sigma^2)^T$ without independence leads to 
\begin{align}
  p(\mu, \sigma^2 ) & \propto (\sigma^2)^{-3/2} \label{eq:jpri}. 
\end{align}
These correspond to $\sigma^2 \sim$ $\Gamma^{-1}(u,v)$, with $u=v=0$ for the
prior given by \eqref{eq:jindpri}, and $u = \frac{1}{2}, v = 0$ for the prior given by
\eqref{eq:jpri}.

\citet[p. 251]{Box:Tiao:92} decompose the likelihood by group means, e.g.\
$\overline{Y}_{i.}$ in \eqref{eq:ydecomposed}, to place a prior directly on
$\sigma^2_{\alpha\epsilon}$.  The joint prior distribution for $\mu,
\sigma_\epsilon^2, \sigma_{\alpha\epsilon}^2$ is then
\begin{align}
  p(\mu, \sigma^2_\epsilon, \sigma^2_{\alpha\epsilon}) \propto
  (\sigma^2_\epsilon \sigma^2_{\alpha\epsilon})^{-1}.
  \label{eq:prior1}
\end{align}
Additionally, Jeffreys' independence prior of the original variance parameters
$(\sigma_\epsilon^2, \sigma_\alpha^2)$ also leads to \eqref{eq:prior1}
\citep{Box:Tiao:92}.  The multivariate analog of this has been used as well by
\citet{Ever:Morr:00a}.  Naturally, a prior may also be placed directly on the
parameter $\sigma_\alpha^2$, although the posterior may no longer be as simple
to work with.

\subsubsection{Conjugacy}
For observations, $Y_i \sim \Norm(\mu, \sigma^2), i = 1, \dots, n$, a
multivariate conjugate prior for the parameter $\btheta = (\mu, \sigma^2)^T$ is
a normal-inverse-gamma distribution, denoted as $\Norm\Gamma^{-1}(\mu_0, \tau, u,
v)$ with $\mu_0 \in \cR, \tau, u, v > 0$.  More specifically,
\begin{align}
  \mu \mid \sigma^2 \sim & \; \Norm( \mu_0, \; \frac{\sigma^2}{\tau} ),\\
  \sigma^2 \sim & \;  \Gamma^{-1}( u, \; v ), 
\end{align}
with joint density,
\begin{align}
  p( \mu, \sigma^2 ) & = p( \mu \vert \sigma^2 ) \cdot p( \sigma^2) \notag \\
& =  (2 \pi \frac{\sigma^2}{\tau}  )^{-1/2} \exp\left( -\frac{\tau }{2 \sigma^2}(\mu -
    \mu_0)^2 \right) \cdot \frac{v^u}{\Gamma(u)}(\sigma^2)^{-u-1} \exp\left(
    -\frac{v}{\sigma^2} \right).  \label{eq:ninvgden}
\end{align}
Priors corresponding to \eqref{eq:jindpri} and
\eqref{eq:jpri} are then denoted by $\Norm\Gamma^{-1}(0,0,-\frac{1}{2},0)$ and by
$\Norm\Gamma^{-1}(0,0,0,0)$, respectively.  Conjugate priors of this form have
been used extensively, although often with precision, $\tau = (\sigma^2)^{-1}$,
resulting in a normal-gamma distribution \cite[p.136]{Bern:Smit:94}.  The
utility of this general parameterization is in being able to conform to
different prior specifications while maintaining conjugacy.  The full model with
likelihood and prior is posterior is
\begin{align}
  Y \mid \mu, \sigma^2 & \sim \Norm(\mu,\, \sigma^2) \label{eq:unidata}, \\
  (\mu, \sigma^2) & \sim \Norm\Gamma^{-1}(\mu_0,\, \tau,\, u,\, v)\label{eq:uniprior}, 
\end{align}
with posterior distribution given by
\begin{align}
  (\mu, \sigma^2) \mid Y & \sim \Norm\Gamma^{-1}\left( \frac{\tau\mu_0 +
    n\overline{y}}{\tau + n},\, 
  \tau + n,\, u+\frac{n}{2},\, v + \frac{1}{2}\left[ \sum_i(y_i - \overline{y})^2 +
  \frac{(\overline{y} - \mu_0)^2}{n^{-1} + \tau^{-1}} \right] \right).  \label{eq:unipost}
\end{align}

\section{Comprehensive ANOVA}\label{sec:comprANOVA}
Following the view of \citet{Gelm:05a} we see the hierarchical Bayesian 
approach towards ANOVA (Section~\ref{sec:ch1multilevel}) as a means to unify the
two distinct fixed and random effects models.  In addition to the hierarchical
model structure a  Bayesian model specification is intuitive and practical.  By
following this approach the challenges discussed in Section~\ref{subsec:issues}
are resolved.  

Hierarchical Bayesian models are typically considered simply as mixed effects
models within the statistical community.  However, because mixed effects models
do not typically provide assessments of uncertainty of the variance component
estimates, nor is variability of the observed set of factor levels examined by
default, we do not see this as truly providing a comprehensive approach towards
ANOVA.  As stated, multilevel modeling seems to be a more natural strategy. As a
result much of the work with multilevel models, including analysis of variance
according to various factors, has largely taken place in other domains,
primarily the social sciences \citep{Gold:95,Gelm:06,Snij:Bosk:11}.  This can be
seen as a failure of the statisticians, as \citet{Hube:11} states, ``the
consequences of not being able to adequately summarize and disseminate common
methodologies may be a divergence of statistics, that each field develops its
own version of statistics.''  By presenting ANOVA in a more general hierarchical
framework we are also, ``unifying the philosophies, concepts, statistical
methods, and computational tools'' \citep{Linds:etal:04}. 

The unification of fixed and random effect models is clearly seen in the
graphical model of Figure~\ref{fig:bgm}.  The successive layers of
distributional assumptions is shown clearly here.  The inner-most box represents
the fixed effects model, while the middle box represents the random effects
model.  The hierarchical Bayesian ANOVA model, or simply \textit{comprehensive
ANOVA}, is then represented by the outer-most box.  The diagram explicitly shows
the unification of the models and immediately conveys a general view of ANOVA to
students and researchers not familiar with variance analyses.  The notation
created by \citet{Eise:47} is here, where ANOVA$_1$ corresponds to $M_1$, the
fixed effects model; and $M_2$ to ANOVA$_2$, the random effects model.  Instead
of $M_3$, which refers to a mixed effects model, we have chosen to allow
ANOVA$_3$ to refer to a fully Bayesian parameterization.  This can be confusing
though, as \cite{Cox:Solo:03} have pointed out, ``Occasionally the word Bayesian
is used for any analysis involving more than one level of random variation.'' We
agree with them, in that this can seem quite confusing, but nonetheless consider
ANOVA$_3$ as a Bayesian analysis of variance procedure.  

\newcommand{\mycirc}{\circle{8}}
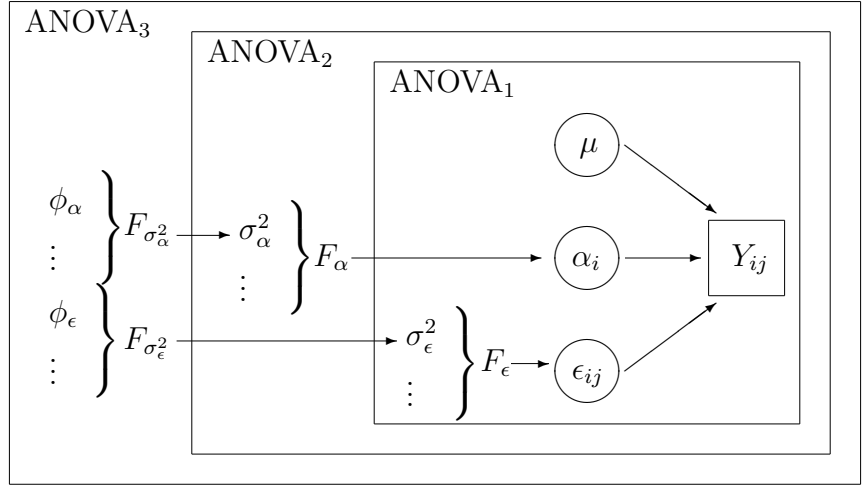
\begin{figure}[t]
  \centering
  \setlength{\unitlength}{1mm}
  \linethickness{0.1mm}
  \begin{picture}(112, 64) 
  \put(48,8){\line(1,0){56}} 
  \put(48,8){\line(0,1){48}} 
  \put(104,8){\line(0,1){48}}
  \put(48,56){\line(1,0){56}}
  \put(50,52){\textrm{ANOVA}$_1$}

  \put(24,4){\line(1,0){84}} 
  \put(24,4){\line(0,1){56}} 
  \put(108,4){\line(0,1){56}}
  \put(24,60){\line(1,0){84}}
  \put(26,56){\textrm{ANOVA}$_2$}

  \put(0,0){\line(1,0){112}} 
  \put(0,0){\line(0,1){64}}
  \put(112,0){\line(0,1){64}}
  \put(0,64){\line(1,0){112}}
  \put(2,60){\textrm{ANOVA}$_3$}

  \put(3,33){$\left. \begin{array}{l} \phi_\alpha \\ \vdots \end{array}\right\}$}
  \put(15,33){$F_{\sigma^2_\alpha}$}
  \put(22,33){\vector(1,0){7}}

  \put(28,29){$\left. \begin{array}{l} \sigma^2_\alpha \\ \vdots \end{array}\right\}$}
  \put(40,29){$F_\alpha$}
  \put(45,30){\vector(1,0){25}}

  \put(3,18){$\left. \begin{array}{l} \phi_\epsilon \\ \vdots \end{array}\right\}$}
  \put(15,18){$F_{\sigma^2_\epsilon}$}
  \put(22,19){\vector(1,0){29}}

  \put(50,15){$\left. \begin{array}{l} \sigma^2_\epsilon \\ \vdots \end{array}\right\}$}
  \put(62,15){$F_\epsilon$}
  \put(66,16){\vector(1,0){5}}

  \put(76,45){\mycirc}
  \put(75,44){$\mu$}
  \put(81,45){\vector(4,-3){12}}

  \put(76,30){\mycirc}
  \put(74,29){$\alpha_i$}
  \put(81,30){\vector(1,0){10}}
  
  \put(76,15){\mycirc}
  \put(74,14){$\epsilon_{ij}$}
  \put(81,15){\vector(4,3){12}}

  \put(92,25){\line(1,0){10}}
  \put(92,25){\line(0,1){10}}
  \put(102,35){\line(0,-1){10}}
  \put(102,35){\line(-1,0){10}}
  \put(95,29){$Y_{ij}$}
  \end{picture}
  \caption{Graphical model representing successive assumptions for
    the fixed effect (inner box), random effect (middle box), and fully Bayesian
    (outer box) specifications.}
  \label{fig:bgm}
\end{figure}


Analysis of variance in this framework allows the questions discussed in the
Introduction to be addressed, and also resolves many of the issues discussed in
Section~\ref{subsec:issues}.   \citet{Gelm:05a} presents graphical summaries of
this ANOVA approach that allow for visual comparison of confidence intervals for
the variance components, which is possible for both finite and superpopulation
variance parameters.  In Table~\ref{tab:banovatab} a proposed alternative to the
traditional ANOVA table is shown.  Commonly significance in the classical ANOVA
table is merely a function of power.  That is, given enough observations, nearly
any effect will be deemed as statistically significant.  Alternatively,
Table~\ref{tab:banovatab} provides estimates of the variance parameters, both
finite and superpopulation, as well as a probabilistic assessment of practical
significance.  This is done with a direct comparison of posterior distributions
of all variance components against the error variance $\sigma^2_\epsilon$.  A
probability regarding hypothesis \eqref{eq:hyp2}, i.e.\ that the superpopulation
variance $\sigma^2_\alpha$ is equal to zero, can be given as well.  This
probability, Pr$(\sigma^2_\alpha = 0 \vert Y)$, thus provides an assessment of
statistical significance.

\begin{table}[b]
\begin{center}
\caption{Comprehensive ANOVA summary utilizing posterior distributions to obtain
a summary of variance parameters (in units of standard deviation).  Quantiles
are used to provide a type of confidence interval.  The probability
$\Pr(\sigma_\alpha>\sigma_\epsilon | Y)$ provides an assessment on practical
significance for the parameter.} \label{tab:banovatab} 
\vspace*{-5mm}
\begin{tabular}{lrrrrr}
  \hline
  &  Parameter & Mean & Median & Uncertainty Interval  & Sig.\ Rel.\ to Errors \\ 
  \hline
  $\alpha$  & (finite) $s_\alpha$        & $\E[ s_\alpha \vert Y]$ & $Q_{0.5}[s_\alpha \vert Y]$ & 
            $\left(Q_{0.025}[ s_\alpha \vert Y],\,Q_{0.975}[ s_\alpha \vert Y]\right)$ & 
            Pr$(s_\alpha > \sigma_\epsilon \vert Y)$ \\
            & (super) $\sigma_\alpha$   & $\E[ \sigma_\alpha \vert Y]$ & $Q_{0.5}[\sigma_\alpha \vert Y]$ & 
            $\left(Q_{0.025}[ \sigma_\alpha \vert Y],\,Q_{0.975}[ \sigma_\alpha \vert Y]\right)$ & 
            Pr$(\sigma_\alpha > \sigma_\epsilon \vert Y)$ \\
 $\epsilon$ & $\sigma_\epsilon$ & $\E[ \sigma_\epsilon \vert Y]$ & $Q_{0.5}[\sigma_\epsilon \vert Y]$ & 
            $\left(Q_{0.025}[ \sigma_\epsilon \vert Y],\,Q_{0.975}[ \sigma_\epsilon \vert Y]\right)$ & 
            $ - $ \\
   \hline
\end{tabular}
\end{center}
\end{table} 

\subsection{One-way Model}
In the case of a single source of variation, as in model
\eqref{eq:1waystandard}, the model can be stated as $Y_{ij} = \alpha_i +
\epsilon_{ij}$.  To illustrate the basic point of view of our ANOVA approach we
first focus on this problem. 

\subsubsection{Model Specification}
A particularly useful parameterization in the one-way configuration, that allows
different variance parameterizations while maintaining conjugacy, is an
extension of the normal-inverse-gamma distribution.  This involves an additional
inverse-gamma distribution for the added variance component.  The resulting
distribution is given by
\begin{align}
  \alpha_i \mid \alpha_0, \tau_\alpha, \tau_\epsilon, \sigma_\alpha^2,
  \sigma_\epsilon^2 \sim & \; \Norm \left( \alpha_0, \; \left[
    \frac{\tau_\alpha}{\sigma_\alpha^2} + 
    \frac{\tau_\epsilon}{\sigma_\epsilon^2} \right]^{-1} \right),\label{eq:ninvginvgA}\\
   \sigma_{\alpha\epsilon}^2 \mid \sigma_\epsilon^2 
    \sim & \; \Gamma^{-1} ( u_\alpha, \; v_\alpha ),\label{eq:ninvginvgB} \\
  \sigma_\epsilon^2 \sim & \; \Gamma^{-1} ( u_\epsilon, \; v_\epsilon ),
  \label{eq:ninvginvgC}
\end{align}
with $i = 1, \dots, n_I$ corresponding to the number of groups, and additional
parameter $\kappa_\epsilon$ such that $\sigma_{\alpha\epsilon}^2 =
\sigma_\alpha^2 + \kappa_\epsilon\sigma_\epsilon^2$.  Noting that
$\sigma^2_{\alpha\epsilon}$ is analogous to the variance of mean of the
observations at an individual factor level, as in Section~\ref{sec:randeff}.

The variance parameters and factor levels can then be jointly specified as a
combination of normal and inverse gamma distributions, i.e.\
$\Norm\Gamma^{-1}\Gamma^{-1}(\alpha_0, \tau_\alpha, \tau_\epsilon,
\kappa_\epsilon, u_\alpha, v_\alpha, u_\epsilon, v_\epsilon)$, with certain
values of hyperparameters, or limits thereof, yielding different prior
specifications such as those discussed earlier.  

This is in general, however, not a conjugate model specification, i.e.\ the
posterior distribution will not be of the same family as the prior distribution.
Although the posterior $\sigma_\epsilon^2$ continues to follow an inverse-gamma
distribution, with density
\begin{align*}
  p(\sigma_\epsilon^2 \mid Y) & \propto (\sigma_\epsilon^2)^{-u_\epsilon -\frac{n-n_I}{2} - 1} 
  \exp\left( -\frac{1}{\sigma_\epsilon^2}\left[ v_\epsilon +
  \frac{1}{2}\sum_{ij}(y_{ij} - \widehat{\alpha}_i)^2 \right] \right),
\end{align*}
the posterior distribution of $\sigma_\alpha^2 +
\kappa_\epsilon\sigma_\epsilon^2$ will not be the same type as its prior for
arbitrary values of hyperparameters $\tau_\epsilon,\kappa_\epsilon$.  More
specifically, the posterior density of $\sigma_\alpha^2$ is
\begin{align*}
  p(\sigma_\alpha^2 \mid Y, \sigma_\epsilon^2) & \propto (\sigma_\alpha^2 +
  \kappa_\epsilon \sigma_\epsilon^2)^{-u_\alpha-1} (\varsigma^2_{\alpha\epsilon} +
  \frac{\sigma_\epsilon^2}{n_J})^{-n_I/2} \\
  & \qquad \times \exp\left( -\frac{v_\alpha}{\sigma_\alpha^2 +
  \kappa_\epsilon \sigma_\epsilon^2} -
  \frac{1}{2} \frac{1}{\varsigma^2_{\alpha\epsilon} +
  \frac{\sigma_\epsilon^2}{n_J}} \sum_i(\widehat{\alpha}_i - \alpha_0)^2 \right),
\end{align*}
where $\varsigma^2_{\alpha\epsilon} = (\frac{\tau_\alpha}{\sigma_\alpha^2} +
\frac{\tau_\epsilon}{\sigma_\epsilon^2})^{-1}$, which can be seen as a type of
shifted inverse-gamma distribution.  However, rather than normalizing the
posterior density so that it is proper when constrained to non-negative values,
it is often more informative to consider a mass point at zero; allowing for the
hypothesis \eqref{eq:hyp2} to be tested (see Section~\ref{sec:examples}). 

Each individual $\alpha_i$ is however normally distributed, with posterior
density
\begin{align*}
  p(\alpha_i \mid Y, \sigma_\epsilon^2, \sigma_\alpha^2) & \propto 
  Q_\alpha^{1/2} \exp \left( -\frac{Q_\alpha}{2}
  \left[\alpha_i - Q_\alpha^{-1} \left(\frac{1}{\varsigma^2_{\alpha\epsilon}}\alpha_0 +
  \frac{n_J}{\sigma_\epsilon^2}\widehat{\alpha}_i\right) \right]^2 \right),
\end{align*}
where $Q_\alpha = \frac{1}{\varsigma^2_{\alpha\epsilon}} +
\frac{n_J}{\sigma_\epsilon^2} = \frac{\tau_\alpha}{\sigma_\alpha^2} +
\frac{\tau_\epsilon + n_J}{\sigma_\epsilon^2}$.  

\subsubsection{Conjugate Prior}
Setting $\tau_\epsilon = 0, \kappa_\epsilon = \frac{\tau_\alpha}{n_J}$, for the
prior and likelihood
\begin{align}
  (\alpha_i, \sigma_\alpha^2, \sigma_\epsilon^2) & \sim \; \Norm\Gamma^{-1}\Gamma^{-1}(\alpha_0, \, 
    \tau_\alpha,\; \tau_\epsilon=0,\; \kappa_\epsilon=\frac{\tau_\alpha}{n_J},\; 
    u_\alpha,\; v_\alpha,\; u_\epsilon,\; v_\epsilon)\label{eq:1waypri}, \\
    Y_{ij} \mid \alpha_i, \sigma_\epsilon^2 & \sim \; \Norm(\alpha_i,\,
    \sigma_\epsilon^2) \label{eq:1waydata}, 
\end{align}
gives way to the posterior distribution
\begin{align}
  (\alpha_i, \sigma_\alpha^2, \sigma_\epsilon^2) \mid Y & \sim \; \Norm\Gamma^{-1}\Gamma^{-1} \left( 
    \left[ \frac{\tau_\alpha}{\sigma_\alpha^2} + \frac{n_J}{\sigma_\epsilon^2} \right]^{-1}
    \left[ \frac{\tau_\alpha}{\sigma_\alpha^2}\alpha_0 +
      \frac{n_J}{\sigma_\epsilon^2}\widehat{\alpha}_i \right], \, 
    \tau_\alpha,\; n_J,\; \frac{\tau_\alpha}{n_J}, \; 
    u_\alpha + \frac{n_I}{2}, \right. \notag \\
  & \left. \qquad \qquad \qquad v_\alpha + \frac{\tau_\alpha}{2}\sum_i(\widehat{\alpha}_i - \alpha_0)^2, \; 
    u_\epsilon + \frac{n-n_I}{2}, \; v_\epsilon + \frac{1}{2}\sum_{ij}(y_{ij} - \widehat{\alpha}_i)^2 
    \right), \label{eq:1waypost}
\end{align}
thus maintaining conjugacy.  Particularly beneficial is that with this
factorization there is no need for MCMC sampling.  Rather, posterior draws can
be taken immediately without burn-in nor thinning.  A single realization from
the joint posterior is found by sampling from $p(\sigma_\epsilon^2 \vert Y)$,
$p(\sigma_\alpha^2 \vert Y, \sigma_\epsilon^2)$, and then $p(\alpha_i \vert Y,
\sigma_\epsilon^2, \sigma_\alpha^2)$ using \eqref{eq:ninvginvgC},
\eqref{eq:ninvginvgB}, \eqref{eq:ninvginvgA} with the parameters updated using
the observations.  

Selecting appropriate values hyperparameters can then be done as follows.  For
invariant priors of variance parameters $u_\alpha = v_\alpha = u_\epsilon =
v_\epsilon = 0$ is used.  To maintain conjugacy $\tau_\epsilon=0,
\kappa_\epsilon = \frac{\tau_\alpha}{n_J}$ is used.  Practical values of
$\tau_\alpha$ and $\alpha_0$ may then be found using an empirical Bayes approach
and yield $\tau_\alpha = 1$ and $\alpha_0 = n_I^{-1}\sum_i\widehat{\alpha}_i$,
i.e.\ the overall mean.  

For more general models, with a mean term, additional factors, interactions,
etc., it is possible to consider several such
normal--inverse-gamma--inverse-gamma distributions, where the single
inverse-gamma distribution of the errors, $\sigma^2_\epsilon$, is common to all.
One may then use a prior distribution for the factor levels under a linear
constraint so that the posterior distributions can also be factored similarly.
This allows not only for conjugacy, but also facilitates computation in a way
that even for models with many factors, samples from the posterior can
efficiently be drawn without the need for MCMC.  This is seen in
\citet{Gein:etal:12a}.

\section{Examples}\label{sec:examples}
\subsection{Rails Data}\label{sec:rails}
For illustration of the various analysis variance methods consider the balanced
one-way design for data consisting of six railway rails \citep{Devo:00,
Pinh:Bate:00}.  Each rail has been measured three times for the amount of time
that it takes a certain type of ultrasonic wave to travel the length of the
rail.  The objective of any initial analysis is most likely to investigate the
(a) variation due to measurement error and (b) variation due to the rails
themselves in terms of both statistical and practical significance.
Additionally, one may be interested predicting travel time for a future
measurement.  This can be considered for either (c) one of these rails as well
as (d) a future rail that has not yet been seen.  For this one-way analysis we
consider the simple cell-means model
\begin{align}
  Y_{ij} = \alpha_i + \epsilon_{ij}, \quad i = 1,\dots,6, \quad j =
  1,2,3.  \label{eq:1waycellmeans}
\end{align}

\subsubsection{Conventional Methods}
For this one-way model error terms $\epsilon_{ij}$ are assumed to be iid
$\Norm(0, \sigma_\epsilon^2)$, and group terms $\alpha_i$ as unknown constants,
so that the observations are
\begin{align}
  Y_{ij} \mid \alpha_i, \sigma_\epsilon^2 & \sim \Norm(\alpha_i,\,
  \sigma_\epsilon^2). 
  \label{eq:m1}
\end{align}
The model assumes that the six rails that have been observed are the only rails
that are of interest.  This is a fixed effects model, which is to say that the
population of rails has been exhausted by the sample.

Questions (a) and (b) can be reasonably addressed using this model, although
purely from a statistically significant point of view.  Results (see
Table~\ref{tab:one}) indicate that the hypothesis \eqref{eq:hyp1} should be
rejected, but are not able to say anything explicitly about the practical
significance of the rails.
\begin{table}[h]
\begin{center}
\caption{One-way ANOVA of Railway Rails}
\begin{tabular}{lrrrrr}
  \hline
 & Df & Sum Sq & Mean Sq & F value & Pr($>$F) \\ 
  \hline
  Rail & 5 & 9310.50 & 1862.10 & 115.18 & 0.000000 \\ 
  Residuals & 12 & 194.00 & 16.17 &  &  \\ 
   \hline
\end{tabular}
\label{tab:one}
\end{center}
\end{table}

Question (c) could be answered by looking at the standard error for the estimate
$\widehat{\alpha}_i$, to obtain an estimate of the expected travel time.
Question (d) can, however, not be answered because of the assumed fixed effect.
To address this question the rails must be considered as a random effect, i.e.\
assumed to come from a greater population of rails.  

\begin{figure}[b]
\begin{center}
  \vspace{3mm}
  \includegraphics{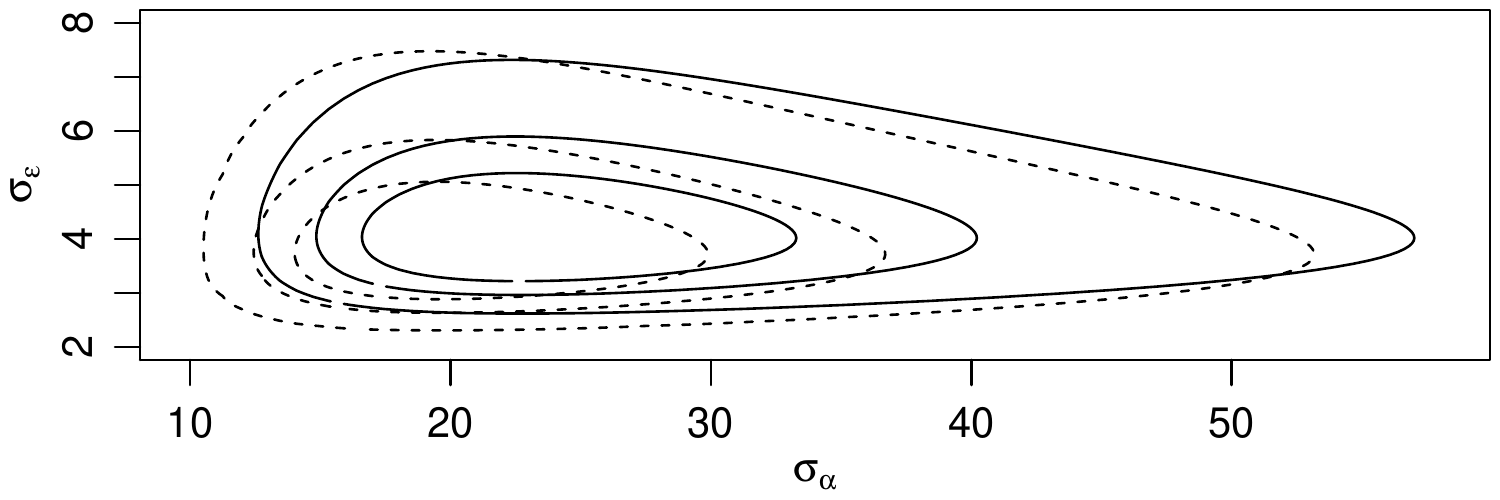}
\end{center}
\vspace*{-8mm}
\caption{Comparison of confidence regions for superpopulation standard
  deviations based on $\chi^2$ approximation of relative log-likelihood (solid)
  and using the highest-posterior-density (dotted).  Contours correspond to
  confidence levels $0.50, 0.75,$ and $0.95$ (small to large).}
\label{fig:ex01:freqvsbanova}
\end{figure}

Alternatively a random effects model may be used, in which the between-rail
variability for a large, potentially infinite (super) population of rails is of
interest.  The hypothesis to be tested is then \eqref{eq:hyp2} and, as discussed
in Section~\ref{sec:randeff}, uses the same F-statistic as for the fixed effects
model.  A more informative summary is often to identify a confidence region of
the variance components, $\sigma^2_\alpha, \sigma^2_\epsilon$, as seen in
Figure~\ref{fig:ex01:freqvsbanova}.  

\subsubsection{Comprehensive ANOVA}
\begin{figure}[h]
\begin{center}
  \includegraphics{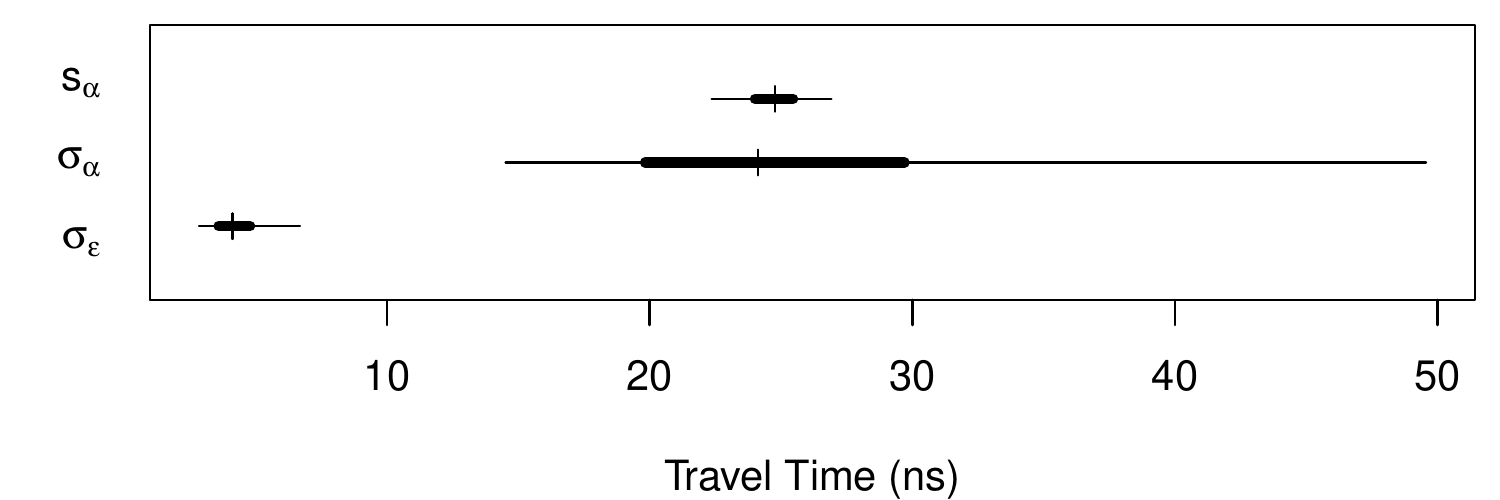}
\end{center}
\vspace*{-8mm}
\caption{Graphical summary of posterior quantiles for variance component 
  (super and finite population) parameters.}
\label{fig:ex1banovaints}
\end{figure}

Analogous to the ANOVA summary provided by Table~\ref{tab:one}, but including
both finite and superpopulation variances, Table~\ref{tab:ex01mybanova}
presents a clearer view on the practical significance of the rails.
Figure~\ref{fig:ex1banovaints} similarly summarizes the analysis.  From the
graphical plot statistical significance is suggested by the fact that the
intervals do not extend to cover $0$.  Because variance parameter must be
nonnegative, it is preferable to assemble a mass point at $0$ when the posterior
has negative support.  This allows for the probability $p(\sigma^2_\alpha = 0
\vert Y)$ to be used to test the hypothesis of $\eqref{eq:hyp2}$, which for this
dataset has probability zero.  
\begin{table}[h]
\begin{center}
\caption{Bayesian ANOVA Table: Posterior distributions are used to obtain
two estimates of the variability.  Quantiles provide an assessment of the
uncertainty in these estimates.  The probability, e.g.\
Pr$(\sigma_\alpha>\sigma_\epsilon)$, provide a relative comparison of each
variance parameter to the measurement variability.}
\vspace{1mm}
\begin{tabular}{lrrrrr}
  \hline
  &  Parameter & Mean & $Q_{0.5}$ & $(Q_{0.025},Q_{0.975})$ & Pr($ > \sigma_\epsilon$) \\ 
  \hline
 Rails  & (finite) $s_\alpha$        & $24.69$ & $24.71$ & $(22.46,26.83)$ & $1.000$ \\
        &  (super) $\sigma_\alpha$   & $25.96$ & $23.89$ & $(14.55,49.20)$ & $1.000$ \\
 Errors &          $\sigma_\epsilon$ & $ 4.27$ & $ 4.10$ & $( 2.87, 6.58)$ & $ - $ \\
   \hline
\end{tabular}
\label{tab:ex01mybanova}
\end{center}
\end{table} 

\subsection{Simulated Data}\label{subsec:simdata}
In the following example a comparison of practical and statistical significance
is illustrated using both classical ANOVA as well as the more comprehensive
Bayesian ANOVA.  Data of the form $Y_{ij} = \alpha_i + \epsilon_{ij}$ is
generated where $\alpha_i \sim \Norm(0,\, \sigma^2_\alpha)$ and $\epsilon_{ij}
\sim \Norm(0,\, \sigma^2_\epsilon = 1)$, for $i = 1,\dots,n_I,\; j = 1,
\dots,n_J$ for a total of $n = n_I\cdot n_J$ observed values.  This is done
under two distinct cases
\begin{center}
\begin{tabular}{l c c }
  \textrm{Case A:} & $\sigma_\alpha^2=\frac{1}{2}$, & $n_J=6$, \\ 
   \textrm{Case B:} & $\sigma_\alpha^2=2$, & $n_J=2$, \\ 
\end{tabular}
\end{center}
and with $n_I = 5$ for both.  Using the conventional ANOVA method
(Table~\ref{tab:ex02classic}) there is not any discernible differences between
the two datasets.  Statistical significance is approximately equivalent because
of the balance of statistical power and the difference in the variance
components $\sigma^2_\alpha$ and $\sigma^2_\epsilon$.  

\begin{figure}[h]
\begin{minipage}[b!]{0.49\linewidth}
\includegraphics[width=\textwidth]{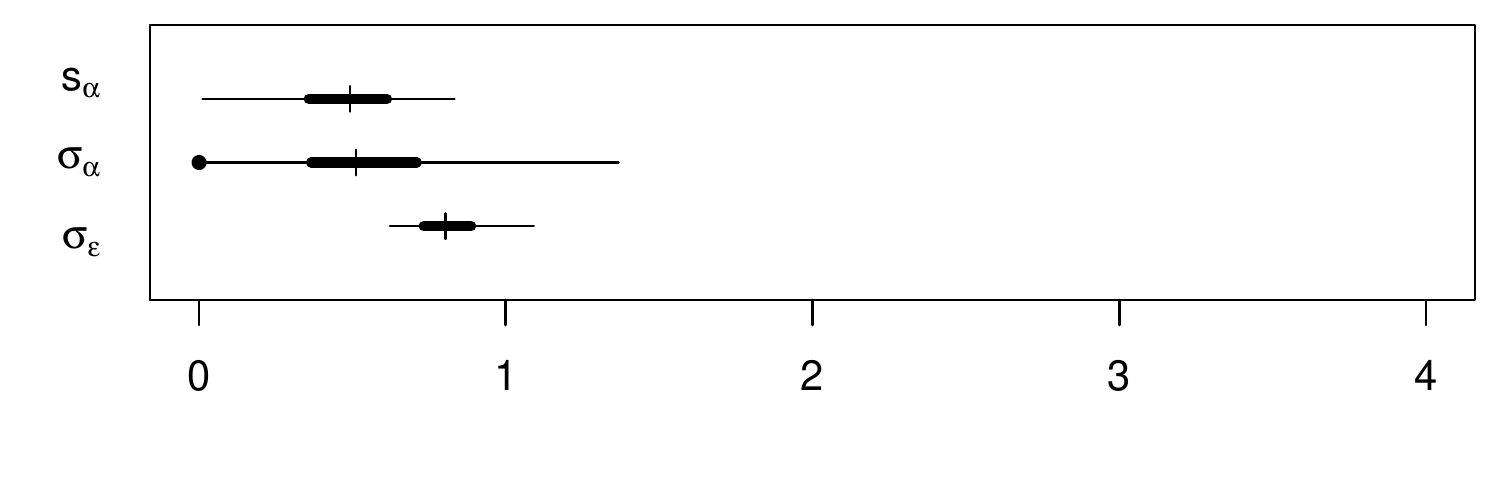}
\end{minipage}
\begin{minipage}[b]{0.02\linewidth}
\hspace{1mm}
\end{minipage}
\begin{minipage}[b!]{0.49\linewidth}
\includegraphics[width=\textwidth]{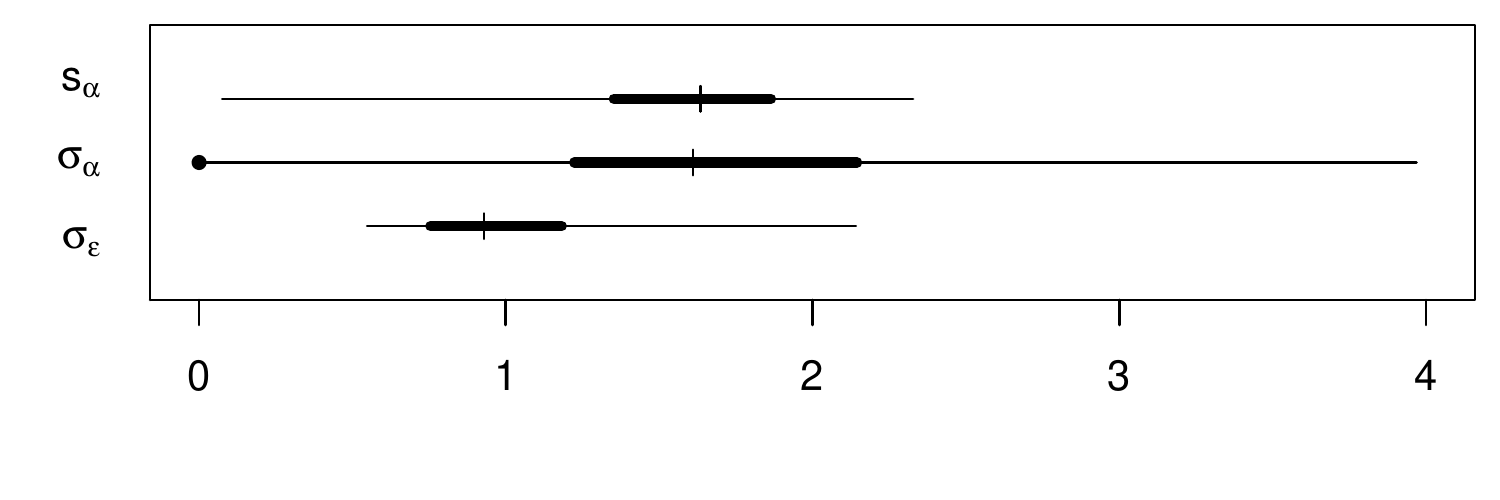}
\end{minipage}
\vspace*{-8mm}
\caption{Posterior uncertainty intervals of variance parameters shown
graphically for both cases.  Thick line segments correspond to $50\%$
uncertainty and thin line segments to $95\%$.  Vertical marks denote the
posterior median.  A point at the end of an interval denotes evidence that the
variance component is zero and can be compared to the corresponding hypothesis
test.}
\label{fig:ex02intervalplots}
\end{figure}

By assembling a mass point at zero whenever the posterior has support for
negative values it possible to use the probability $p(\sigma^2_\alpha = 0 \vert
Y)$ to test the hypothesis of $\eqref{eq:hyp2}$.  Interestingly, this posterior
probability is $0.0263$ for case A and $0.0258$ for case B, values which are
comparable to their corresponding $p$-values in Table~\ref{tab:ex02classic}.
This is also noted by the end-points corresponding to the $0.025$ level of
uncertainty for the intervals shown in Figure~\ref{fig:ex02intervalplots}.  

A more informative and comprehensive summary of the data is provided by
the Bayesian ANOVA table (Table~\ref{tab:ex02newtable}). This provides not only
estimates of the variance components, but also an indication of the practical 
significance of the factor $\alpha$ when observed with error $\epsilon$. 

\begin{table}[h]
\caption{Classical ANOVA table to summarize the decomposition of variance.  Case
  A (left), with $n_I = 5,n_J = 6, \sigma^2_\alpha = \frac{1}{2}$, represents low
practical significance of factor $\alpha$. Case B (right), with $n_I = 5,n_J =
2, \sigma^2_\alpha = 2$, represents strong practical significance.  Despite
practical differences between the two cases, $p$-values are nearly equal.}
\vspace{1mm}
\begin{minipage}[t!]{0.49\linewidth}\centering
\scriptsize
\begin{center}
\begin{tabular}{lrrrrr}
\hline
& Df & Sum Sq & Mean Sq & F value & Pr($>$F) \\ 
\hline
$\alpha$ & 5 & 9.70 & 1.94 & 3.08 & 0.0267 \\ 
$\epsilon$ & 25 & 15.75 & 0.63 &  &  \\ 
\hline
\end{tabular}
\end{center}
\end{minipage}
\begin{minipage}[b]{0.02\linewidth}
\hspace{1mm}
\end{minipage}
\begin{minipage}[t!]{0.49\linewidth}\centering
\scriptsize
\begin{center}
\begin{tabular}{lrrrrr}
\hline
& Df & Sum Sq & Mean Sq & F value & Pr($>$F) \\ 
\hline
$\alpha$ & 5 & 27.69 & 5.54 & 7.31 & 0.0239 \\ 
$\epsilon$ & 5 & 3.79 & 0.76 &  &  \\ 
\hline
\end{tabular}
\label{tab:ex02classic}
\end{center}
\end{minipage}
\end{table}

\begin{table}[h]
\caption{Bayesian ANOVA tables to summarize the variance decomposition.  Case A
(left) and Case B (right) illustrate a situation in which factor $\alpha$ has
low, or high practical significance.}
\vspace{1mm}
\begin{minipage}[t!]{0.49\linewidth}\centering
\scriptsize
\begin{center}
\begin{tabular}{lrrrrr}
  \hline
  &  Parm & Mean & $Q_{0.5}$ & $(Q_{0.025},Q_{0.975})$ & Pr($ > \sigma_\epsilon$) \\ 
  \hline
 $\alpha$  & $s_\alpha$         & $0.48$ & $0.49$ & $(0.02, 0.84)$ & $0.07$ \\
        &  $\sigma_\alpha$      & $0.57$ & $0.51$ & $(0.02, 1.34)$ & $0.19$ \\
 $\epsilon$ & $\sigma_\epsilon$ & $0.81$ & $0.80$ & $(0.62, 1.07)$ & $ -  $ \\
   \hline
\end{tabular}
\end{center}
\end{minipage}
\begin{minipage}[b]{0.02\linewidth}
\hspace{1mm}
\end{minipage}
\begin{minipage}[t!]{0.49\linewidth}\centering
\scriptsize
\begin{center}
\begin{tabular}{lrrrrr}
  \hline
  &  Parm & Mean & $Q_{0.5}$ & $(Q_{0.025},Q_{0.975})$ & Pr($ > \sigma_\epsilon$) \\ 
  \hline
 $\alpha$  & $s_\alpha$         & $1.59$ & $1.66$ & $(0.12, 2.30)$ & $0.84$ \\
           & $\sigma_\alpha$    & $1.74$ & $1.61$ & $(0.11, 3.84)$ & $0.82$ \\
 $\epsilon$ & $\sigma_\epsilon$ & $1.04$ & $0.92$ & $(0.52, 2.30)$ & $ -  $ \\
   \hline
\end{tabular}
\end{center}
\end{minipage}
\label{tab:ex02newtable}
\end{table}

\vspace{5mm}

\section{Discussion}
The major contribution of this paper can be seen as ideological in nature, in
that the standard method of analysis of variance is treated as a useful
procedure to practitioners.  As a result, although rigorous treatment is given,
the discussion has been restricted to relatively simple designs.  Extending from
a one-way balanced ANOVA to many factors can be considered as a trivial step
from here.  However, much more work is needed in order for the method to be able
to be widely accepted.  Issues of unbalanced designs, non-orthogonal predictors,
and generalized linear models are all necessary for the widespread usage of any
statistical method.  Therefore these are all issues that are to be examined in
greater detail in the future.

\bibliographystyle{thesiswiley}
\bibliography{../../Dropbox/UZH/research/thesis_bib}

\begin{thebibliography}{}

\bibitem[Bates and DebRoy, 2004]{Bate:DebR:04}
Bates, D. and DebRoy, S. (2004).
\newblock Linear mixed models and penalized least squares.
\newblock {\em Journal of Multivariate Analysis}, {\bf91},  1--17.

\bibitem[Bernardo and Smith, 2000]{Bern:Smit:94}
Bernardo, J.~M. and Smith, A. (2000).
\newblock {\em Bayesian Theory (Wiley Series in Probability and Statistics)}.
\newblock John Wiley and Sons Ltd.

\bibitem[Box and Tiao, 1992]{Box:Tiao:92}
Box, G. E.~P. and Tiao, G.~C. (1992).
\newblock {\em Bayesian Inference in Statistical Analysis}.
\newblock Wiley-Interscience.

\bibitem[Cochran, 1934]{Coch:34}
Cochran, W.~G. (1934).
\newblock The distribution of quadratic forms in a normal system, with
  applications to the analysis of covariance.
\newblock {\em Mathematical Proceedings of the Cambridge Philosophical
  Society}, {\bf30},  178--191.

\bibitem[Cox and Solomon, 2003]{Cox:Solo:03}
Cox, D. and Solomon, P.~J. (2003).
\newblock {\em Components of Variance}, volume~97 of {\em Monographs on
  statistics and applied probability}.
\newblock Chapman and Hall.

\bibitem[Devore, 2000]{Devo:00}
Devore, J. (2000).
\newblock {\em Probability and Statistics for Engineering and the Sciences}.
\newblock Brooks/Cole, Thomson Learning.

\bibitem[Eisenhart, 1947]{Eise:47}
Eisenhart, C. (1947).
\newblock The assumptions underlying the analysis of variance.
\newblock {\em Biometrics}, {\bf3},  pp. 1--21.

\bibitem[Everson and Morris, 2000]{Ever:Morr:00a}
Everson, P.~J. and Morris, C.~N. (2000).
\newblock Inference for multivariate normal hierarchical models.
\newblock {\em Journal of the Royal Statistical Society. Series B (Statistical
  Methodology)}, {\bf62},  pp. 399--412.

\bibitem[Fidler {\it et~al.}, 2004]{Fidl:etal:04}
Fidler, F., Geoff, C., Mark, B., and Neil, T. (2004).
\newblock Statistical reform in medicine, psychology and ecology.
\newblock {\em Journal of Socio-Economics}, {\bf33},  615--630.

\bibitem[Fisher, 1925]{Fish:25}
Fisher, R. (1925).
\newblock {\em Statistical methods for research workers}.
\newblock Edinburgh Oliver \& Boyd, 1 edition.

\bibitem[Gardner and Altman, 1986]{Gard:Altm:86}
Gardner, M. and Altman, D. (1986).
\newblock Confidence intervals rather than p values: estimation rather than
  hypothesis testing.
\newblock {\em British medical journal (Clinical research ed.)}, {\bf292},
  746.

\bibitem[Geinitz {\it et~al.}, 2012]{Gein:etal:12a}
Geinitz, S., Furrer, R., and Sain, S.~R. (2012).
\newblock {Multivariate analysis of global climate projections via rank
  deficient Bayesian ANOVA}.
\newblock {\em Journal of the Royal Statistical Society: Series C (under
  revision for resubmission)}.

\bibitem[Gelman, 2005]{Gelm:05a}
Gelman, A. (2005).
\newblock Analysis of variance: Why it is more important than ever.
\newblock {\em Annals of Statistics}, {\bf33},  1--31.

\bibitem[Gelman and Hill, 2006]{Gelm:06}
Gelman, A. and Hill, J. (2006).
\newblock {\em Data Analysis Using Regression and Multilevel/Hierarchical
  Models}.
\newblock Cambridge University Press, 1 edition.

\bibitem[Goldstein, 1995]{Gold:95}
Goldstein, H. (1995).
\newblock {\em Multilevel Statistical Models}.
\newblock Kendall's Library of Statistics. E. Arnold.

\bibitem[Greenland, 2000]{Gree:00}
Greenland, S. (2000).
\newblock Principles of multilevel modelling.
\newblock {\em International Journal of Epidemiology}, {\bf29},  158--167.

\bibitem[Hald, 1998]{Hald:98}
Hald, A. (1998).
\newblock {\em A history of mathematical statistics from 1750 to 1930}.
\newblock Wiley Series in Probability and Statistics. Wiley.

\bibitem[Hoaglin {\it et~al.}, 1991]{Hoag:Most:Tuke:91}
Hoaglin, D., Mosteller, F., and Tukey, J. (1991).
\newblock {\em Fundamentals of Exploratory Analysis of Variance}.
\newblock Wiley series in probability and mathematical statistics: Applied
  probability and statistics. Wiley.

\bibitem[Huber, 2011]{Hube:11}
Huber, P. (2011).
\newblock {\em Data Analysis: What Can Be Learned From the Past 50 Years}.
\newblock Wiley Series in Probability and Statistics. John Wiley \& Sons.

\bibitem[Ioannidis, 2005]{Ioan:05}
Ioannidis, J. P.~A. (2005).
\newblock Why most published research findings are false.
\newblock {\em Public Library of Science Medicine}, {\bf2},  e124.

\bibitem[Jeffreys, 1946]{Jeff:46}
Jeffreys, H. (1946).
\newblock An invariant form for the prior probability in estimation problems.
\newblock {\em Proceedings of the Royal Society of London. Series A,
  Mathematical and Physical Sciences}, {\bf186},  pp. 453--461.

\bibitem[Kreft {\it et~al.}, 1998]{Kref:deLe:98}
Kreft, I., De~Leeuw, J., and de~Leeuw, J. (1998).
\newblock {\em Introducing multilevel modeling}.
\newblock Sage London.

\bibitem[Leinonen {\it et~al.}, 2008]{Lein:08}
Leinonen, T., O'Hara, R.~B., Cano, J.~M., and Meril{\"a}, J. (2008).
\newblock Comparative studies of quantitative trait and neutral marker
  divergence: a meta-analysis.
\newblock {\em J. Evolutionary Bio.}, {\bf21},  1--17.

\bibitem[Lencina {\it et~al.}, 2005]{Lenc:etal:05}
Lencina, V.~B., Singer, J.~M., and Stanek, E.~J. (2005).
\newblock Much ado about nothing: the mixed models controversy revisited.
\newblock {\em International Statistical Review}, {\bf73},  9--20.

\bibitem[Lindsay {\it et~al.}, 2004]{Linds:etal:04}
Lindsay, B.~G., Kettenring, J., and Siegmund, D. (2004).
\newblock {A Report on the Future of Statistics}.
\newblock {\em Statistical Science}, {\bf19},  387--413.

\bibitem[Lunn {\it et~al.}, 2000]{Lunn:etal:00}
Lunn, D.~J., Thomas, A., Best, N., and Spiegelhalter, D. (2000).
\newblock Winbugs - a bayesian modelling framework: Concepts, structure, and
  extensibility.
\newblock {\em Statistics and Computing}, {\bf10},  325--337.

\bibitem[Nakagawa and Cuthill, 2007]{Naka:Cuth:07}
Nakagawa, S. and Cuthill, I. (2007).
\newblock Effect size, confidence interval and statistical significance: a
  practical guide for biologists.
\newblock {\em Biological Reviews}, {\bf82},  591--605.

\bibitem[Nelder, 1999]{Neld:99}
Nelder, J.~A. (1999).
\newblock From statistics to statistical science.
\newblock {\em J. R. Statist. Soc. D}, {\bf48},  257--269.

\bibitem[Nelder, 2008]{Neld:08}
Nelder, J.~A. (2008).
\newblock What is the mixed-models controversy?
\newblock {\em International Statistical Review}, {\bf76},  134--135.

\bibitem[Pinheiro and Bates, 2000]{Pinh:Bate:00}
Pinheiro, J. and Bates, D. (2000).
\newblock {\em Mixed-Effects Models in S and S-Plus}.
\newblock Statistics and Computing. Springer.

\bibitem[Pinheiro {\it et~al.}, 2006]{Pinh:etal:06}
Pinheiro, J., Bates, D., DebRoy, S., and Sarkar, D. (2006).
\newblock {\em nlme: Linear and nonlinear mixed effects models}.

\bibitem[Qian and Shen, 2007]{Qian:Shen:07}
Qian, S.~S. and Shen, Z. (2007).
\newblock Ecological applications of multilevel analysis of variance.
\newblock {\em Ecology}, {\bf88},  2489--2495.

\bibitem[Rao, 1997]{Rao:97}
Rao, P. (1997).
\newblock {\em Variance components: mixed models, methodologies and
  applications}, volume~78.
\newblock Chapman \& Hall/CRC.

\bibitem[Sain {\it et~al.}, 2011]{Sain:etal:11}
Sain, S.~R., Nychka, D.~W., and Mearns, L. (2011).
\newblock {Functional ANOVA and regional climate experiments: a statistical
  analysis of dynamic downscaling}.
\newblock {\em Environmetrics}, {\bf22},  700--711.

\bibitem[Savage, 1957]{Sava:57}
Savage, I.~R. (1957).
\newblock {Nonparametric Statistics}.
\newblock {\em Journal of the American Statistical Association}, {\bf52},
  331--344.

\bibitem[Searle {\it et~al.}, 1992]{Sear:etal:92}
Searle, S., Casella, G., and McCulloch, C. (1992).
\newblock {\em Variance components}.
\newblock Wiley series in probability and mathematical statistics: Applied
  probability and statistics. Wiley.

\bibitem[Snijders and Bosker, 2011]{Snij:Bosk:11}
Snijders, T. and Bosker, R. (2011).
\newblock {\em Multilevel analysis: An introduction to basic and advanced
  multilevel modeling}.
\newblock Sage Publications Limited.

\bibitem[Von~Korff {\it et~al.}, 1992]{VonK:etal:92}
Von~Korff, M., Koepsell, T., Curry, S., and Diehr, P. (1992).
\newblock Multi-level analysis in epidemiologic research on health behaviors
  and outcomes.
\newblock {\em American Journal of Epidemiology}, {\bf135},  1077--1082.

\bibitem[Voss, 1999]{Voss:99}
Voss, D.~T. (1999).
\newblock Resolving the mixed models controversy.
\newblock {\em American Statistician}, {\bf53},  352--356.

\bibitem[Yoccoz, 1991]{Yocc:91}
Yoccoz, N. (1991).
\newblock Use, overuse, and misuse of significance tests in evolutionary
  biology and ecology.
\newblock {\em Bulletin of the Ecological Society of America}, {\bf72},
  106--111.

\end{thebibliography}

\end{document}